# Permanent Magnet Penning Trap

Daniel C. Barnes[1, 3, a] and Daniel R. Knapp[2,3]

[1] Coronado Consulting, Lamy, NM 87540, USA
[2] Medical University of South Carolina, Charleston, SC 29425, USA, and
[3] Wilhelm Bratwurst Institute, Charleston, SC 29407, USA

**ABSTRACT:** The Penning trap has been investigated as the basis of a small nuclear fusion reactor using a superconducting solenoid magnet. To extend this investigation, we designed, constructed, and evaluated a permanent magnet Penning trap. The device consists of a solenoid formed from an annular array of neodymium bar magnets between two iron pole pieces designed to give a uniform magnetic field in the central volume of the device. Critical to achieving the uniform solenoidal field is an iron "equatorial ring" supported within the annular array of magnets. A nonmagnetic titanium Penning trap with hyperbolic surfaces designed to produce a spherical potential well was mounted inside the permanent magnet assembly. The trap was fitted with a nonmagnetic hairpin filament electron source and demonstrated to produce electron trapping at the theoretically predicted magnetic fields and trap potentials. Trap potentials achievable were limited by electrical breakdown within the trap operating in constant potential mode. Efforts were made to extend the trap potentials using pulsed anode voltages, but nuclear fusion in a Penning trap has not yet been demonstrated. The design and construction of the permanent magnet solenoid and nonmagnetic trap are presented here as potentially useful also in other studies.

[a] Corresponding author: email coronadocon@msn.com

# I. INTRODUCTION

## A. Motivation

The Penning trap[1] was investigated at Los Alamos National Laboratory (LANL) in the late nineties as the basis of a possible nuclear fusion reactor [2—7]. Unfortunately, the work was terminated before this approach could be fully investigated. The LANL work utilized a superconducting solenoid magnetic field. We sought to re-examine the Penning trap as a small fusion reactor with a more compact and less expensive magnetic field source. This paper describes the design, construction, and operation of the permanent magnet Penning trap constructed for this work. Efforts to achieve nuclear fusion by this approach are ongoing, but the permanent magnet Penning trap might also prove useful in other applications.

## B. Background

The present system was designed to advance the concept of Penning fusion.[2] In this approach, electrons are spherically focused to the center of a hyperbolic Penning trap tuned to produce a spherical well in the effective potential. The over-convergence of primary, injected electrons then produces a central virtual cathode which may be used to trap ions electrostatically near the trap center. These ions can reach thermonuclear energies if the applied potential and corresponding magnetic field is sufficiently high, and fusion reactions among the ions can be observed to produce neutrons and other fusion products in exothermic reactions.

Electrons produced on the axis produce a Brillouin flow[8] (zero canonical $P_\theta$). and experience an effective electron potential in a Penning trap with electrostatic potential of a single $P_2$ spherical harmonic produced by hyperbolic electrodes. The effective potential is given by

$$V_{eff} = \frac{eV_A}{3a^2}(2z^2 - r^2) + \frac{e^2 B^2}{8m} r^2 \qquad (1)$$

where $V_A$ is the applied anode-cathode voltage, $a$ the reference spherical radius for this potential, $e$ and $m$ the electron charge and mass, and $r$ and $z$ the usual cylindrical coordinates, and where we have set the effective potential to zero at the spherical origin where $r = 0 = z$. The choice

$$V_A = \frac{eB^2 a^2}{8m} \qquad (2)$$

causes the effective potential to become that of a spherical harmonic well, leading to the central focusing of injected electrons and formation of the desired virtual cathode.

The earlier LANL work[3—4] demonstrated focusing of electrons in a liquid-He temperature system when the applied fields very nearly satisfy the resonance condition of Eq. (2). In this embodiment, essentially no neutrals or ions could be introduced into the system because of the strong cryo-pumping at such low temperatures. Some work was also done in a room-temperature system, but results were limited.

The objectives of the present work were to achieve low field errors (both magnetic and electrostatic) in a Ultra-High Vacuum (UHV) room-temperature system which would allow both the required potentials and low-pressure gas fill to realize such a configuration. Challenges with this approach include:

- Production and maintenance of a magnetic field with good uniformity
- Production of the required electrostatic configuration
- Access for electron injection and collection
- Resistance to unwanted electrical discharges associated with the High Voltage (HV)

Some of these challenges have been addressed and successful techniques demonstrated, while others, particularly reliable operation at very high applied voltages, remain to be resolved in the future.

## C. Present outline

This paper describes the approaches taken and results obtained. The next section describes the design and construction of the experimental apparatus. Experimental results are presented in section III. Section IV summarizes and presents conclusions.

## II. DESIGN AND CONSTRUCTION

The overall design of the permanent magnet Penning trap consists of a 10 mm internal radius hyperbolic Penning trap mounted inside a permanent magnet solenoid, with the entire assembly fitted inside a high vacuum system. The details of the design and construction of the experimental system are the following.

### A. Vacuum System

The vacuum system main chamber is built from a ten inch Conflat (CF1000) nipple mounted vertically with four additional side ports (two each CF600 and CF275) added to form a six way cross. The top CF1000 flange serves as the main access for insertion and removal of the permanent magnet assembly and Penning trap using an overhead electric winch. The side ports accommodate a multiconductor low voltage feedthrough, a 30 kV high voltage feedthrough, a viewport, and additional vacuum connections. A Varian TV-301 250 l/s turbo pump is attached to the bottom CF1000 via an ISO100 gate valve. A tee connector between the turbo pump and gate valve has an NW25 port to which is connected a flexible stainless steel (bellows) hose to allow fine control of chamber pumping via a bellows valve with the main gate valve closed. This pumping capability is used for fine pressure control when the chamber is backfilled with a gas (e.g. deuterium) for experiments.

The additional vacuum connections via the CF275 side port include the aforementioned differential pumping connection, connection to a Granville-Phillips VQM830 residual gas analyzer (RGA) with Granville-Phillips Micro Ion ATM gauge (which can also be differentially pumped via a bellows valve), a connecting tube between the chamber and RGA for sampling the chamber atmosphere, and connecting tube for internal attachment of a gas feed line to the Penning trap. The main chamber pressure is monitored by a Bayard-Alpert gauge tube attached to the top chamber flange.
The magnet and trap assembly is supported at the vertical center of the chamber by a stainless steel stand inside the chamber. The vacuum system is capable of reaching pressures in the mid $10^{-8}$ Torr region (ca. $6 \times 10^{-6}$ Pa). Fill gases were supplied from a buffer volume via a Granville-Phillips 203 Variable Leak valve.

### B. Permanent magnet solenoid

A permanent magnet system was designed and constructed with the goals of: a) producing a uniform field; b) providing axial access for electron injection; c) flexibility of field strength; and d) relative ease of assembly/disassembly.

The desired uniformity was set at a variation of less than $10^{-3}$ relative over the spherical trap volume of a 1 cm radius sphere. This uniformity was chosen to be consistent with expected tolerances of machining and alignment. This specified fraction of 1 cm amounts to only 10 $\mu$m or approximately 0.0004", which was smaller than the anticipated tolerances of the remaining components.

To provide flexibility and ease of access, the design was chosen as an assembly of two disk-shaped pole pieces of magnetic iron, with an annular array of permanent magnet rods connecting them and providing the required magnetization. This arrangement provides a simple and rigid assembly which was maintained by the magnetic attraction of the poles toward one another and the magnetizing rods. The field strength was varied by inserting more or fewer rods in a roughly uniform arrangement around the annulus.

The shape of the poles was determined to produce the specified uniformity by considering the iron to be infinitely permeable, so that the field external to the poles was the analog of the electrostatic field produced by maintaining the poles at a specified $\pm V$. (Infinite permeability implies that the external magnetic field and proportional induction are strictly normal to the pole surfaces so the external vacuum field is proportional to the electrostatic field just described.) An in-house electrostatic solver, based on a boundary-integral method was used to compute the magnetic fields produced by candidate shapes, and the candidates were chosen for simplicity of machining to be shapes which could be produced easily on a conventional lathe.

The design was based on the intuition that: a) the shape of the surfaces facing away from the trap were not important for field uniformity; b) the shape of the surfaces facing toward the trap should be concave to compensate for the fringing of the field produced by the finite pole radius; and c) access holes on axis would be required through the disk and some compensation of the resulting field perturbation would be desired. A range of shapes were considered, and it was found that desirable configuration was one that consisted of: a single step in thickness producing a concave recess in the poles plus a small boss near the access holes compensating for the field error introduced by the holes plus a flat back surface

This reduced the search space to that of a few parameters, as shown in Fig. 1. The pole radius was chosen as large as practical with constraints of material and fitting into the vacuum system to be 95 mm. The central hole was chosen as small as practical to allow access with a radius of 5 mm. The maximum thickness was chosen to be consistent with a 1" stock at 24.5 mm. The material was Carpenter Consumet® Electrical Iron (Carpenter Technology Corp., Philadelphia, PA), a Si/V magnetic iron alloy with high permeability and excellent vacuum properties.

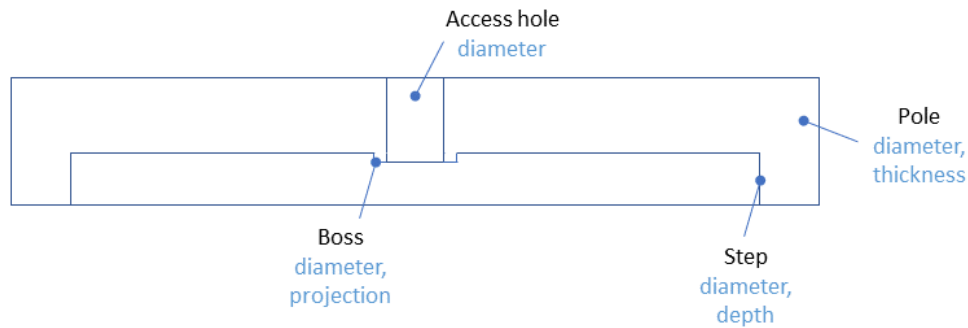

**Figure 1:** Schematic of pole shape, shown as upper pole, with the design parameters indicated.

The remaining four dimensions (step diameter and depth, and boss diameter and projection) were varied to obtain best field uniformity, using the boundary-integral code. The best field uniformity so obtained was still around 1% relative, and it was then discovered that adding a small equatorial ring of magnetic iron could improve this considerably, so the final design was as shown in Fig. 2. The blue-shaded region of Fig. 2a is the annulus where the PM rods were placed. Although the design allowed for up to 1" diameter rods (or squares) ½" [Grade N42 Neodynium (NIB) magnets, Applied Magnets, Plano, TX] magnet rods 2" in length were chosen for convenience and availability.

The field uniformity predicted by the electrostatic calculation is shown in Fig. 3 and is predicted to by $< 2 \times 10^{-4}$ relative over the central 1 cm radius sphere. These predictions were verified by comparison with a model using the COMSOL® Multiphysics finite element approach.

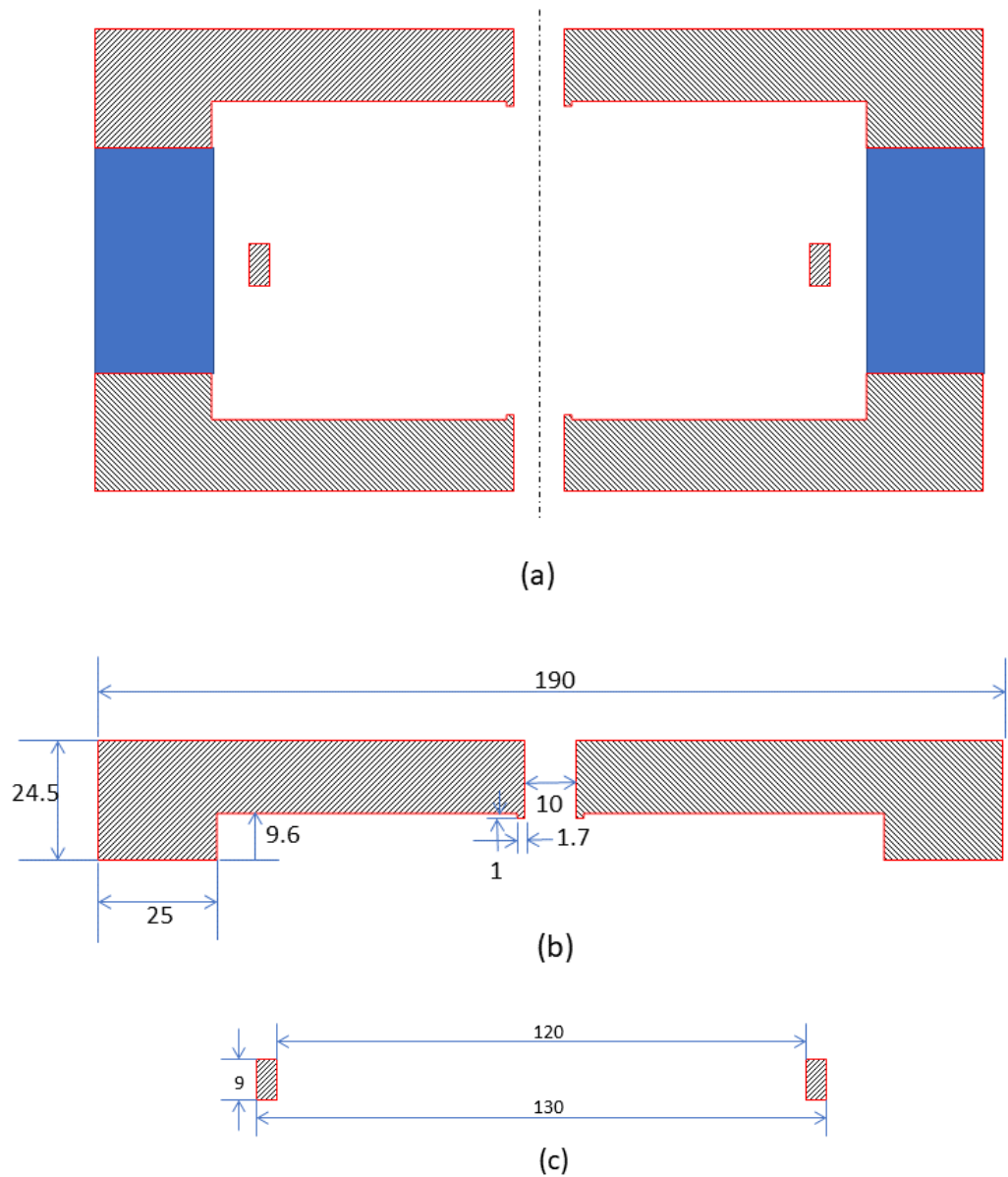

**Figure 2:** Final magnet design: (a) overall arrangement; (b) pole piece with dimensions in mm; (c) equatorial ring with dimensions in mm.

The assembly was completed by non-magnetic (aluminum) pieces which provided alignment of the rods at the lower pole and supported the equatorial ring in its desired mid-plane position. Figure 4 shows the magnet assembly with the trap (see following sub-section) inside before the upper pole is lowered to complete the assembly. The lowering is done with a custom fixture chucked in a milling machine, so that alignment and safe assembly are assured. The field uniformity was confirmed to the accuracy of gaussmeter (model GM-2, AlphaLab Inc., Salt Lake city, UT) measurements, which was unfortunately larger than the design tolerance.

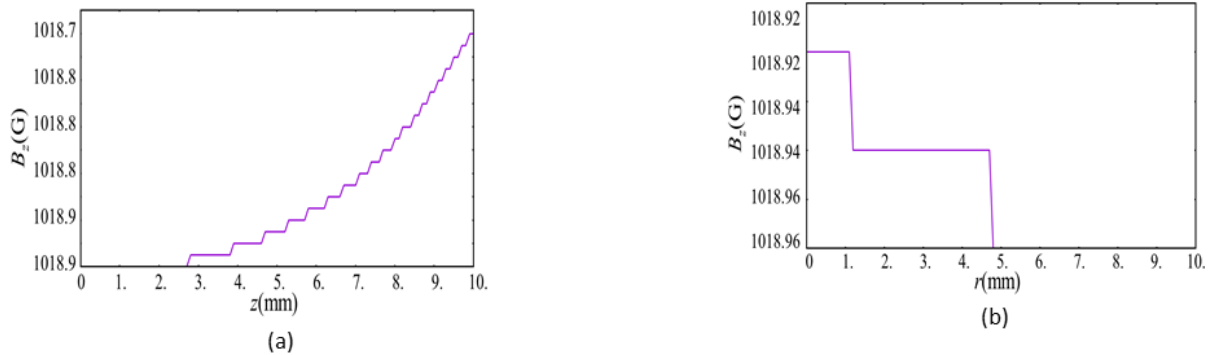

**Figure 3:** Calculated field strength *vs.* distance from trap center: (a) along axis; (b) along radius. Steps are limits of plotting resolution.

Because the unavoidable air gaps between the rods and the poles were not completely controllable, and because the commercially-available NIB rod magnets were not completely characterized as to magnetization, the final field strength was not completely predictable and was instead measured after each assembly, using the gaussmeter. The resulting field strength was largely proportional to the number of rods.

C. **Penning trap**

The Penning trap electrodes shown in Figure 5 were constructed from grade 2 titanium by computer numerical control machining at the Clemson University College of Engineering Machine Shop. The trap was designed with hyperbolic electrode surfaces to produce a spherical potential well according to the standard formulas where $Z_0$ is the vertical distance from the trap center to the surface centers of the upper and lower cathodes and $R_0$ is the radius from the trap center to the nearest surface of the ring anode:

$$r = \sqrt{2(z^2 - Z_0^2)}, Z_0 = 10.0 \tag{3}$$

$$z = \sqrt{\frac{r^2 - R_0^2}{2}}, R_0 = 10.1 \tag{4}$$

The trap was designed as a spherical trap as shown with $R_0$ increased by 1% to prevent injected electrons from touching the anode when the trap is tuned to the spherical point [Eq. (2)]. The cathodes and anode were supported and maintained at the proper spacing using alumina rods and spacer rings sold as spare parts for ion trap mass spectrometers by Scientific Instrument Services, Ringoes, NJ (part numbers Z28 and Z29).

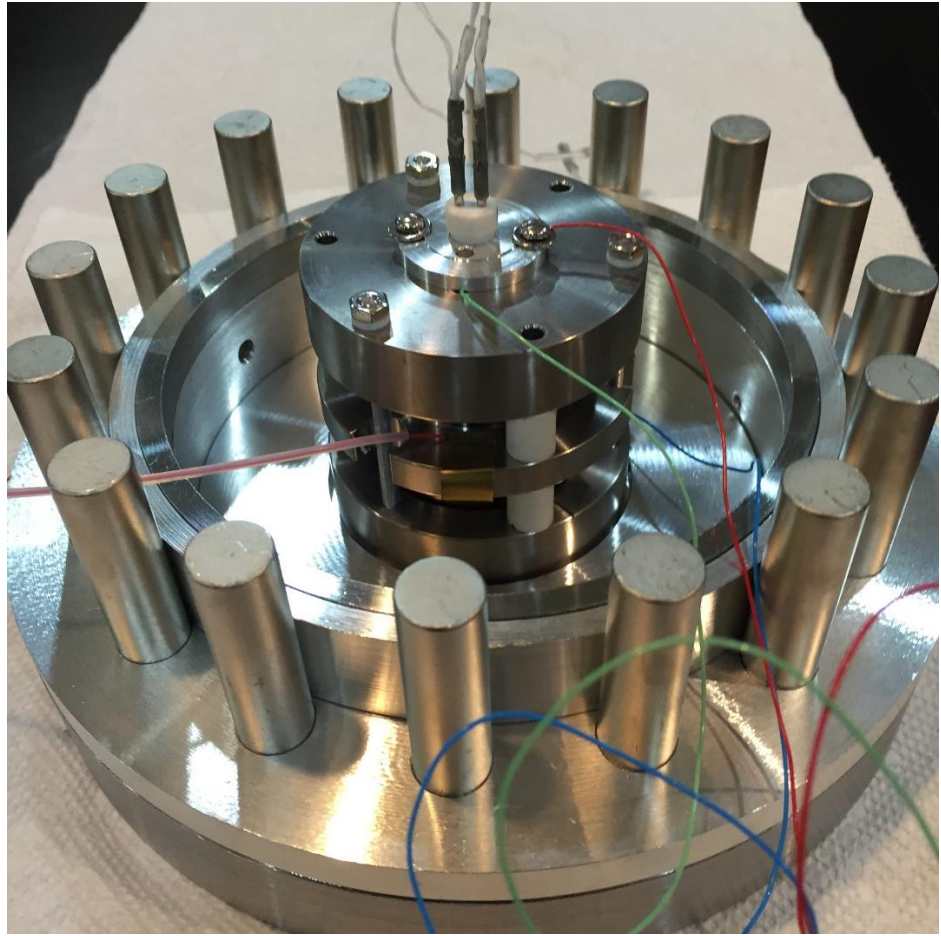

**Figure 4:** Completed magnet assembly with 16 rods inserted.

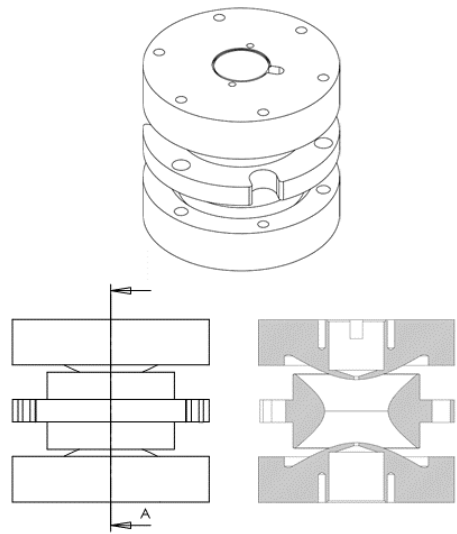

**Figure 5:** Schematic drawing of Penning trap.

D. **Electron source**

Electrons were supplied by a 0.003" tungsten wire "hairpin" filament of the shape commonly used as electron microscope filaments. The ductile tungsten wire used was obtained as model 18306 needle cleaning wires ("type 218 tungsten") from Hamilton Company, Reno, NV, and bent on an aluminum fixture to achieve a reproducible shape. The filament was spot welded to type 316 stainless steel posts mounted in a machined Macor base using "solder glass" (type EGO4000VEG, Ferro Corp., Cleveland, OH). The standard electron microscope filaments initially used had Kovar pins, which were found by modeling (Comsol Multiphysics) to cause significant distortion of the magnetic field uniformity within the trap. The filament was mounted inside a titanium "Wehnelt" cap with a 1.50 mm diameter aperture, which itself was mounted in the upper cathode with alumina ring insulators. The filament center was carefully aligned in the aperture center using a video microscope. The filaments were typically operated at about 2 V and 1.5 A to yield electron emission currents of tens of $\mu$A.

E. **Electronics and diagnostics**

For constant anode potential experiments, the anode potential was supplied by a Bertan 2554-2 0 to ±30 kV, 0-400 microamp power supply module driven by an in-house constructed analog controller that provided the 0 -10 VDC voltage control signal and other interfacing. Anode current was observed via the power supply current monitor signal for plotting anode current vs. anode potential curves.

For pulsed experiments, the Penning trap anode potential was supplied by a Matsusada AKP-40P320 0 - 40 kV power supply that was switched with a 50 kV fast high voltage push pull switch (model HTS-510-10-GSM, Behlke High voltage, Billerica, MA). The power supply was connected to the high voltage push pull switch via a filter consisting of a 10 k$\Omega$ high voltage resistor followed by a 2000 pF high voltage capacitor, which was then connected to the positive terminal of the Behlke switch via a 100 $\Omega$ high voltage resistor. A second 100 $\Omega$ high voltage resistor connected the negative terminal of the Behlke switch to ground (the 100 $\Omega$ resistors are recommended by the switch manufacturer to suppress ringing). Pulses were generated using an Agilent 33220A function generator.

The voltage pulses were observed with a North Star PVM-5 1000:1 high voltage probe attached to a Tektronix TDS-2012C digital storage oscilloscope. Current pulses were observed with Pearson model 2877 current monitor on the high voltage lead and connected to the same oscilloscope via a 1000X battery powered preamplifier attached directly to the Pearson monitor. Preamplifiers using a variety of instrumentation amplifiers IC's were examined (see discussion). Filament power, filament bias, and Wehnelt potential were supplied by separate variable DC power supplies. Magnetic field strengths were measured with the gaussmeter. For the efforts to effect nuclear fusion in the trap, bubble dosimeters (model BD-PMD, Bubble Technology Industries, Chalk River ON, CA) were used as neutron detectors.

III. **EXPERIMENTS**

A. **Steady anode potential experiments**

A series of experiments were carried out at different field strengths achieved by varying the number of magnets. In a typical experiment at a given field strength, a few tens of microamperes of electron current is injected into the trap while the anode potential is gradually increased. The anode current slowly increases with anode potential and shows a sharp peak at the electron trapping potential ("resonance" potential) given by Eq. (2).

This peak is the result of maximum scattering of the electron beam to the anode by the trapped electron cloud at the "resonance." Figure 6 plots a typical data set for a single magnetic field value (0.037 T in this case). Further increases in anode potential beyond the decrease following the resonance peak resulted in a rapidly rising anode current leading to full discharge within the trap. Each increase in field strength required prolonged pumping and gradual surface conditioning to enable scanning through and observing the resonance peak while avoiding discharge.

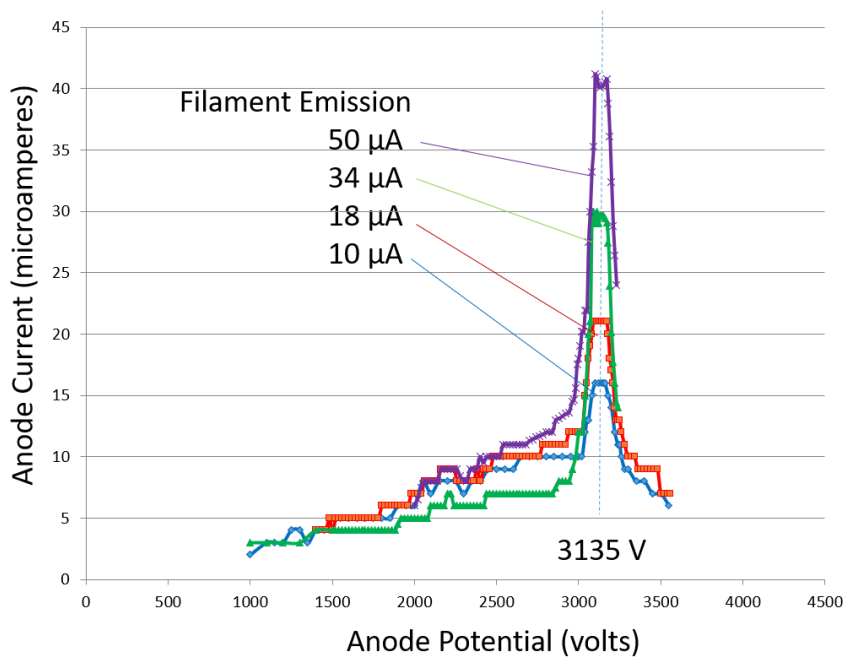

**Figure 6:** Anode current vs. voltage for magnetic field of 0.037 T and several filament emission currents.

Figure 7 is a plot of the observed resonance peaks along with a plot of the predicted resonance potential showing good agreement between theoretical and observed performance. Observation of the peak at 25 kV was difficult due to discharge breakdown in the trap, and this peak was only observed once.

In order to go to higher anode potentials, efforts were made to suppress the discharge in the trap. A series of insulators made from fluoropolymers and polyimides did not solve the problem. Since it was suspected that secondary electron generation contributed to the discharge breakdown, efforts were made to suppress such generation by coating the cathode surfaces with colloidal graphite and with titanium nitride. Neither coating made a significant difference in raising the breakdown potential.

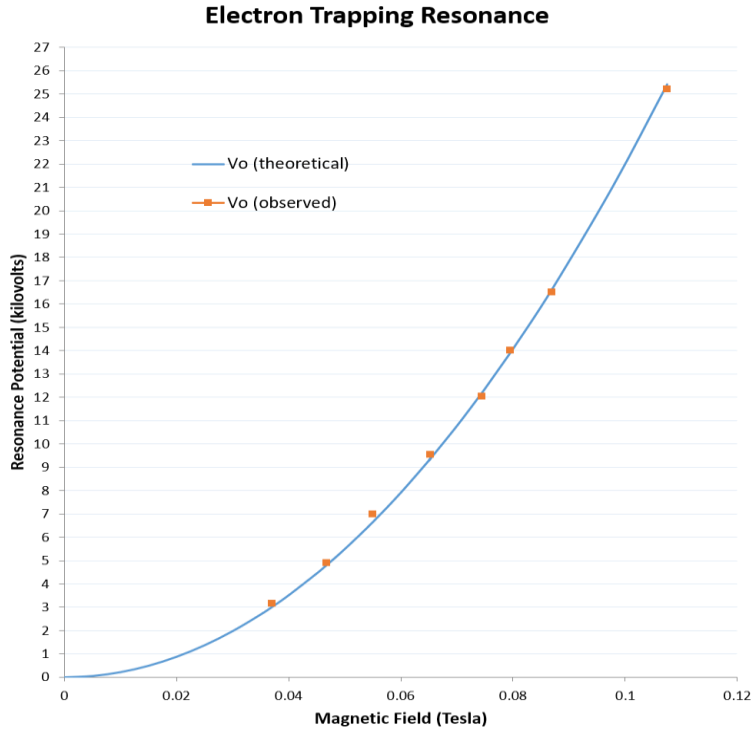

**Figure 7:** Observed resonance peaks *vs.* anode potential along with the predicted theoretical curve [Eq. (2)] for resonance potentials.

### B. Pulsed anode potential experiments

Having reached a barrier in useable constant anode potentials, efforts were made to go to higher potentials by pulsing the high voltage to the anode using a Behlke high voltage push pull switch. The potential pulse shape was monitored with a high voltage oscilloscope probe and showed the expected pulse shapes for pulses examined up to about 25 kV with pulse widths from a few microseconds to 200 microseconds. Observations of the current pulses carried out with a 1 volt per ampere Pearson current transformer using a 1000X preamplifier proved more problematic, with a persistent problem of destroying the amplifier integrated circuits in spite of having diode protection on the amplifier inputs. The discharge behavior of the trap under pulsed conditions was erratic, but tended to improve somewhat with extended pumping and surface conditioning at each increment of voltage. At times the expected current pulse behavior was observed where the breakdown current would set in late in the voltage pulse. In these cases, shortening the pulse did eliminate the breakdown current pulse, but reproducible results could not be achieved. The end result was that we have thus far been unable to usefully extend the anode potential range in pulsed operation to fusion-relevant potentials.

### IV. DISCUSSION AND CONCLUSIONS

A permanent-magnet Penning trap has been designed, constructed, and operated in a room-temperature UHV system. The magnet design successfully produces a uniform field with a non-uniformity of $< 10^{-3}$ relative with a mechanical arrangement which allows relativly easy access and flexibile choice of field strength.

A completely non-magnetic, hyperbolic-electrode Penning trap has been designed, constructed (principally from titanium), and operated at voltages up to 25 kV applied anode potential within this magnet system. Previously reported electron focusing results were confirmed over a range of magnetic fields from 0.037 T to just over 0.1 T and associated resonant anode potentials from 3 kV to just over 25 kV.

Attempts were made to introduce deuterium ions into the system using a static gas fill. While the gas system operated as desired, increased electrical breakdown of the trap interfered with desired control of ion fueling and heating in the central cathode expected at the electron focus. No neutron production was observed in these experiments.

The limiting constraint on operation of such a system, in attempts to push toward fusion conditions in the central focus, is electrical breakdown of the trap. While careful conditioning and maintenance of excellent vacuum quality allowed obtaining the electron focusing results presented previously here, the conditions required to avoid breakdown continue to be studied, with the goal of understanding the interaction of HV, both DC and pulsed, low pressure gas fill, and magnetic field in producing breakdowns. Optimistically, this increased understanding may lead to finding operational regimes which allow higher voltage operation with trapped ions and ultimately demonstration of nuclear fusion.

## V. **ACKNOWLEDGEMENTS**

The authors would like to thank Mike Cassidy and Ben Longmeier of Apollo Fusion for helpful discussion and for the loan of equipment.

16. Daniel R. Knapp and Daniel C. Barnes, "Permanent Magnet Spherical Penning trap as a Small Fusion Source," Summer School on Neutron Detectors and Related Applications 2016 NDRA, June 29-July 2, 2016, Riva del Garda, Italy.

17. Daniel R. Knapp and Daniel C. Barnes, "Permanent Magnet Spherical Penning trap as a Small Fusion Source," 23$^{nd}$ IAEA Technical Meeting on Research Using Small Fusion Devices, March 29-31, 2017, Santiago, Chile.

18. Daniel R. Knapp and Daniel C. Barnes, "Permanent Magnet Spherical Penning trap as a Small Fusion Source," 54th Culham Plasma Physics Summer School, July 17-28, 2017, Abington, UK.

19. Daniel C. Barnes and Daniel R. Knapp, "Spherical Penning Trap as a Small Fusion Source," 20thUS-Japan IEC Workshop, November 8, 2016, University of Wisconsin, Madison, WI.